\documentclass[aps,prb,floatfix,amsmath,amssymb,eqsecnum,nofootinbib,twocolumn,superscriptaddress]{revtex4-1}
\usepackage[usenames,dvipsnames]{color}

\usepackage{graphicx,xcolor}
\usepackage[abs]{overpic}
\usepackage{bm}
\usepackage{amsmath}  
\usepackage[caption=false]{subfig}
\usepackage{relsize}  
\usepackage{amssymb}
\usepackage{graphics}
\usepackage{esint}

\begin{document}
\global\long\def\bra#1{\left\langle #1\right|}
\global\long\def\ket#1{\left|#1\right\rangle }
\global\long\def\ketbra#1#2{\left|#1\vphantom{#2}\right\rangle \left\langle #2\vphantom{#1}\right|}
\global\long\def\braket#1#2{\left\langle #1|#2\right\rangle }
\global\long\def\parenopen#1{\left(#1\right.}
\global\long\def\parenclose#1{\left.#1\right)}
\global\long\def\commopen#1{\left[#1\right.}
\global\long\def\commclose#1{\left.#1\right]}

\title{Thermalization in Open Quantum Systems}

\author{Israel Reichental}
\affiliation{Department of Physics, Technion, Haifa 32000, Israel}

\author{Anat Klempner}
\affiliation{Department of Physics, Technion, Haifa 32000, Israel}

\author{Yariv Kafri}
\affiliation{Department of Physics, Technion, Haifa 32000, Israel}

\author{Daniel Podolsky}
\affiliation{Department of Physics, Technion, Haifa 32000, Israel}
\affiliation{ITAMP, Harvard-Smithsonian Center for Astrophysics, Cambridge, Massachusetts 02138, USA}

\date{\today}

\begin{abstract}
We study thermalization in open quantum systems using the Lindblad formalism. A method that both thermalizes and couples to Lindblad operators only at edges of the system is introduced. Our method leads to a Gibbs state of the system, satisfies fluctuation-dissipation relations, and applies both to integrable and non-integrable systems. Possible applications of the method include the study of systems coupled locally to multiple reservoirs. Our analysis highlights the limits of applicability of the Lindblad approach to study strongly driven systems. 
\end{abstract}

\maketitle

\section{Introduction}

One of the basic problems in many-body physics is conduction between two baths, which act as heat and particle reservoirs.  The problem has been addressed using a host of approaches, for both classical and quantum systems \cite{Landauer1957,Landauer1970,Buttiker1986,Bonetto2000,Lepri2003,Michel2006,Meir1992,Jauho1994}.  More recently, much work has been done using quantum Markov processes based on the Lindblad approach\cite{Lindblad1976,Gorini1976}. These give a host of novel effects, surprisingly even for non-interacting systems.  These include bounds on ballistic transport in the XXZ chain \cite{Prosen}, phase transitions in non-interacting chains held far from equilibrium \cite{Prosena}, and a resonant structure in the thermopower generated between two edges with different temperatures \cite{Dubi}.

Broadly, one can identify two general approaches for using Lindblad operators to drive quantum systems. In the first, the reservoir are modeled by Lindblad operators acting globally on the whole system \cite{Dubi,Saito,Saito2003,Weiss2006,Dolcini2013,Viyuela2012}. The advantages of this approach are that it can be microscopically justified from the derivation of the Lindblad equation \cite{Spohn,Davies1974}, in the limit of a weak coupling between the system and the reservoir; and that, when the system is coupled to a single reservoir, the system equilibrates. It is important to keep in mind that in this approach fluctuation-dissipation relations are violated if the coupling to the reservoir is pushed beyond the weak coupling limit \cite{Keeling}. In the second approach, the reservoir are modeled by Lindblad operators acting locally on a few sites at the edges of the system \cite{Prosen,Prosena,Prosen2011,Wichterich2007,Karevski2013,Weiss2006,Michel2003}.   This approach has the advantage that the non-unitary effect of the reservoir acts only at the edges, and not in the bulk of the system.   However, it is not clear if the local approach can actually model an equilibrated reservoir, namely one that would equilibrate the system when no other reservoirs are present \cite{Znidaric}.  This may be responsible for apparent violations of the second law of thermodynamics, as discussed in Ref. \onlinecite{0295-5075-107-2-20004}. It would be useful to have a method that bridges these two approaches.

In this paper we introduce such a method. Within our approach the reservoir is coupled to the system on one edge, thermalizes it and satisfies fluctuation-dissipation relations. To achieve this,   we model the reservoir as a non-interacting lead together with a bath, which acts on the lead through Lindblad operators.  The lead is, in turn, coupled to the system at a single site. Under certain conditions that we specify, both integrable and non-integrable systems thermalize. The method provides a first step toward using the Lindblad approach to study non-equilibrium situations where the system is coupled to several reservoirs at different points. It also clarifies when the global approach can applied.  A similar approach was introduced in Refs. \onlinecite{Guimaraes2016, Zanoci2016, Mahajan2016} to study thermalization in integrable systems.

Our results highlight the limit of applicability of the Lindblad formalism to study driven system. We show that it works only in a weak coupling limit. In particular, this approach does not allow the study of large currents which do not occur in this limit.

\section{The Model}
\label{sec:model}

\begin{figure}
\centering
\begin{overpic}[width = 0.48\textwidth]{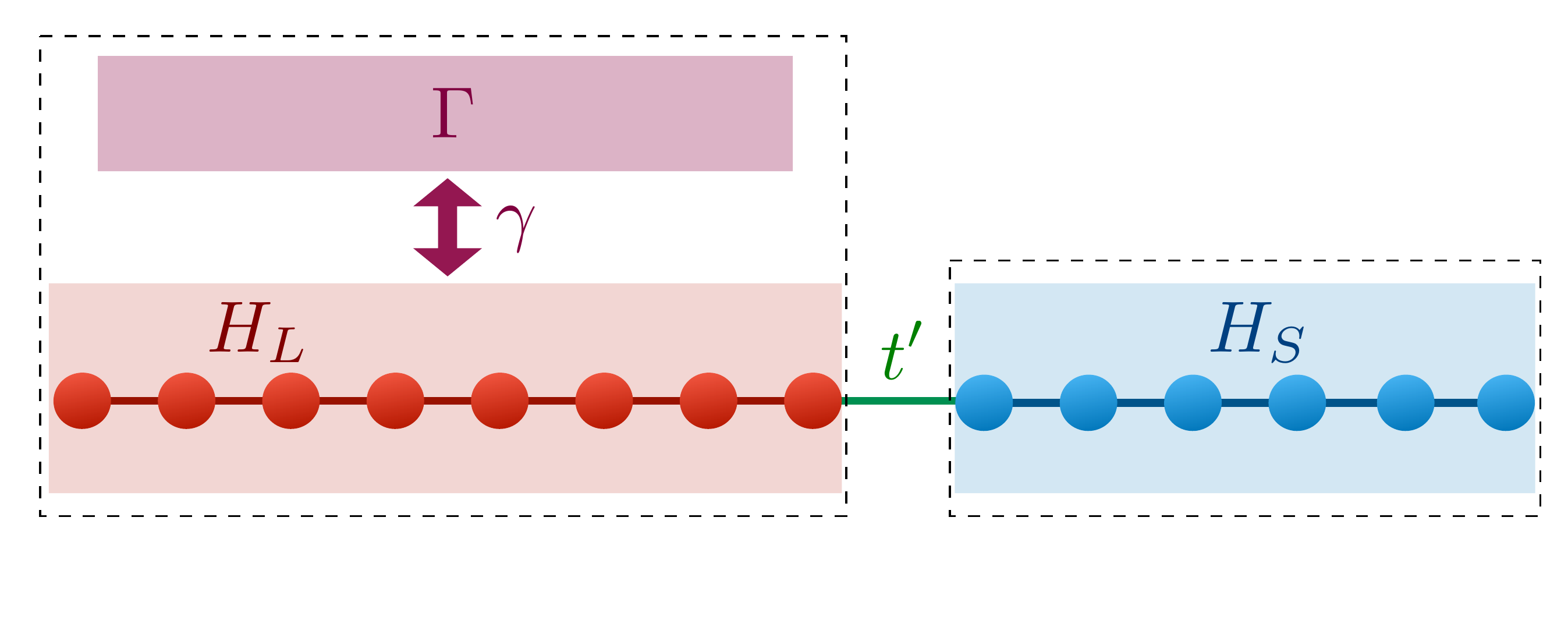}
\put(41,1){\large bath+lead} 
\put(177,1){\large system} 
\end{overpic}
\caption{Illustration of the model.  The reservoir consists of a bath, modeled by Lindblad operators $\Gamma$, acting on a non-interacting lead. For simplicity, we use periodic boundary conditions in the lead.}
\label{fig:physical_setup}
\end{figure}

We consider the setup depicted in Fig.~\ref{fig:physical_setup}.   A one dimensional quantum system, described by a Hamiltonian $H_S$, is coupled to a reservoir through a single link on one of its edges.  The reservoir is composed of a non-interacting one-dimensional lead, described by a Hamiltonian $H_L$, and a bath, which is modeled by a dissipative generator $\hat{\Gamma}$.  The evolution of the total density matrix $\rho$ of the system and lead is then given by 
\begin{equation}
\partial_{t}\rho=-i\left[H_L+H_S+H_{\rm int},\rho\right]+\hat{\Gamma}\rho
\end{equation}
where $H_{\rm int}$ is the coupling between the system and the lead. The dissipative generator is given by
\begin{equation}
\hat{\Gamma}\rho=\frac{1}{2}\sum_{\nu}\gamma_{\nu}\left(2\Gamma_{\nu}\rho\Gamma_{\nu}^{\dagger}-\rho\Gamma_{\nu}^{\dagger}\Gamma_{\nu}-\Gamma_{\nu}^{\dagger}\Gamma_{\nu}\rho\right),\label{eq:dissipative}
\end{equation}
where $\Gamma_\nu$ are quantum jump operators, which act only on the sites of the lead, and $\gamma_{\nu}$ are rates with dimensions of energy.  These are chosen such that when $H_{\rm int}=0$, the lead reaches thermal equilibrium with a Boltzmann measure with inverse temperature $\beta$ and chemical potential $\mu$.

We consider a lead Hamiltonian of the form,
\begin{eqnarray}
H_{L} & = & -t_{L}\sum_{j=1}^{M}\left(b_{j}^{\dagger}b_{j+1}+b_{j+1}^{\dagger}b_{j}\right),
\end{eqnarray}
where $b_{j}$, $b_{j}^{\dagger}$ are annihilation and creation operators for a spinless fermion on site $j$ of the lead, which has $M$ sites.  For simplicity, we consider periodic boundary conditions in the lead, such that $b_{M+1}=b_{1}$.  Then, $H_{L}$ can be diagonalized in Fourier space, by introducing
\begin{eqnarray}
b_m=\frac{1}{\sqrt{M}} \sum_j b_j e^{i 2\pi m j}
\end{eqnarray}
which yields
\begin{eqnarray}
H_{L} & = & \sum_{m}\epsilon_{m}^{L}b_{m}^{\dagger}b_{m}
\end{eqnarray}
where $\epsilon_{m}^L=-2t_L\, \cos (2\pi m)$.

To ensure that the lead thermalizes whenever  $H_{\rm int}=0$, we impose detailed balance with respect to the lead eigenstates $m$.  For each $m$, we introduce two jump operators: $\Gamma_{m-}=b_m$ removes a fermion with rate $\gamma_{m-}=\gamma_m (1-f_m)$, while $\Gamma_{m+}=b_m^\dagger$ injects a fermion with rate $\gamma_{m+}=\gamma_m f_m$.  Here,  $f_{m}= f_{{\rm FD}}\left(\epsilon_{m}^{L}\right)$ and
\begin{equation}
f_{{\rm FD}}\left(\epsilon\right)\equiv\frac{1}{1+e^{\beta\left(\epsilon-\mu\right)}}
\end{equation}
is the Fermi-Dirac distribution.  Then,
\begin{equation}
\frac{\gamma_{m+}}{\gamma_{m-}}=\frac{f_m}{1-f_m}=e^{-\beta(\epsilon_m^L-\mu)},
\end{equation}
and Eq. (\ref{eq:dissipative}) becomes
\begin{eqnarray}
\hat{\Gamma}\rho & = & \frac{1}{2}\sum_{m}\gamma_{m}\commopen{f_{m}\left(b_{m}^{\dagger}\rho b_{m}-b_{m}b_{m}^{\dagger}\rho-\rho b_{m}b_{m}^{\dagger}\right)}\nonumber \\
 &  & +\commclose{\left(1-f_{m}\right)\left(b_{m}\rho b_{m}^{\dagger}-b_{m}^{\dagger}b_{m}\rho-\rho b_{m}^{\dagger}b_{m}\right)}.\label{eq:lindblad_thermalization}
\end{eqnarray}
Throughout, unless otherwise stated, we will concentrate on the case $\mu=0$.

In what follows, we will consider separately both integrable and non-integrable system Hamiltonians $H_S$.  In the integrable case,
\begin{eqnarray}
H_{S} & = & -t_{S}\sum_{j=1}^{N-1}\left(c_{j}^{\dagger}c_{j+1}+c_{j+1}^{\dagger}c_{j}\right),\label{eq:HSint}
\end{eqnarray}
where $c_{j}$, $c_{j}^{\dagger}$ are annihilation and creation operators for a spinless fermion on site $j$ of the system, which has $N$ sites.  In this case one can express $H_{S}$ in terms of the one-particle energy eigenstates,
\begin{eqnarray}
H_{S} & = & \sum_{n}\epsilon_{n}^{S}c_{n}^{\dagger}c_{n},\label{eq:hamiltonian_system_eigenmodes}
\end{eqnarray}
where $\epsilon_{n}^{S}$ is the energy of the $n^{th}$ eigenstates. In the non-integrable case,
\begin{eqnarray}
H_{S} & = & -t_{S}\sum_{j=1}^{N-1}\left(c_{j}^{\dagger}c_{j+1}+c_{j+1}^{\dagger}c_{j}\right)\nonumber\\
&\,&+U \sum_{j=1}^{N-1} n_jn_{j+1}+V \sum_{j=1}^{N-2} n_jn_{j+2}.\label{eq:HSnon_int}
\end{eqnarray}
where $n_j=c_j^\dagger c_j$ is the number of fermions in site $j$.  Throughout, we set $t_{S}=1$.

Finally, we take the coupling between the system and lead to be:
\begin{equation}
H_{\rm int}=-t^{\prime}\left(b_{M}^{\dagger}c_{1}+c_{1}^{\dagger}b_{M}\right).\label{eq:hamiltonian_coulping}
\end{equation}

While the model defined above stands as a phenomenological model on its own, one could imagine deriving it from an underlying microscopic model.  For instance, the dissipative terms could be obtained by coupling the lead sites to an environment by a weak link of strength $t''$.  Then, deriving the Redfield equation for this composite system and integrating out the additional chain yields a Lindblad equation with rates $\gamma_m$ of the order of $\left(t''\right)^{2}D$ where $D$ is density of states of the environment
 \cite{Breuer2002}.  Furthermore, if $t'\ll t''$, then the ensuing dissipative terms would couple globally to the lead (but not the system) in such a way as to bring it into thermal equilibrium.  Therefore, if the dissipative term were to arise from this microscopic point of view, one would require $t'\ll t''$ \cite{Davis1998,Wichterich2007,Purkayastha2016}. 

\section{Integrable Case}

We note that, in the integrable case, the entire Lindblad equation is quadratic in fermionic operators.  This makes it amenable to analytic approaches, as described below.

\subsection{One Site Lead ($M=1$)}

The first case that we investigate is an integrable system with a one site lead, with total Hamiltonian
\begin{eqnarray}
H & = & -t^{\prime}\left(c_{1}^{\dagger}b+b^\dagger c_{1}\right)\nonumber \\
 & & -t_S\sum_{i=1}^{N-1}\left(c_{i}^{\dagger}c_{i+1}+c_{i+1}^{\dagger}c_{i}\right)\label{eq:one_site_hamiltonian}
\end{eqnarray}
and dissipative generator,
\begin{eqnarray}
\hat{\Gamma}\rho & = & \frac{1}{2}\gamma\commopen{f_{0}\left(2b^{\dagger}\rho b-bb^{\dagger}\rho-bb^{\dagger}\rho\right)}\nonumber \\
 &  & +\commclose{\left(1-f_{0}\right)\left(2b\rho b^{\dagger}-b^{\dagger}b\rho-b^{\dagger}b\rho\right)},\label{eq:one_site_lindblad}
\end{eqnarray}
with $f_{0}=\frac{1}{1+e^{-\beta\epsilon_0}}$. This corresponds to a lead Hamiltonian, $H_L=\epsilon_0 b^\dagger b$ (this differs from the convention above, since one cannot define a tight binding model on one site). As we show in Appendix \ref{appen:onesite}, independently of $\beta$, the steady state corresponds to an infinite temperature state. Namely, the reduced density matrix of the system (after tracing out the lead) becomes, in the energy basis,
\begin{eqnarray}
\rho_S=\prod_n \left(f_0 c_n^\dagger c_n+(1-f_0) c_n c_n^\dagger\right).
\end{eqnarray}
This result holds even for more general tight-binding models.

It is natural to ask whether this result is an artifact of the single-site lead, and whether thermal equilibrium is achieved by considering
a more realistic lead composed of multiple sites.  As we will show next, while the situation does improve when the lead is enlarged, thermalization is not achieved even in the infinite lead limit.

\subsection{Multiple Site Lead}

We proceed by extending the setup above to a lead with multiple sites.  In the integrable case, the Lindblad equation is quadratic in fermionic operators, and hence can be solved using a third quantization method \cite{Prosen2008}.  This method involves the diagonalization of a $4(N+M)$-dimensional matrix, allowing us to study fairly large leads and systems numerically.  A detailed description of the third quantization method applied to this class of systems is given in Ref. \onlinecite{Prosen2008}.

\begin{figure}
\centering
\begin{overpic}[width=0.45\textwidth]{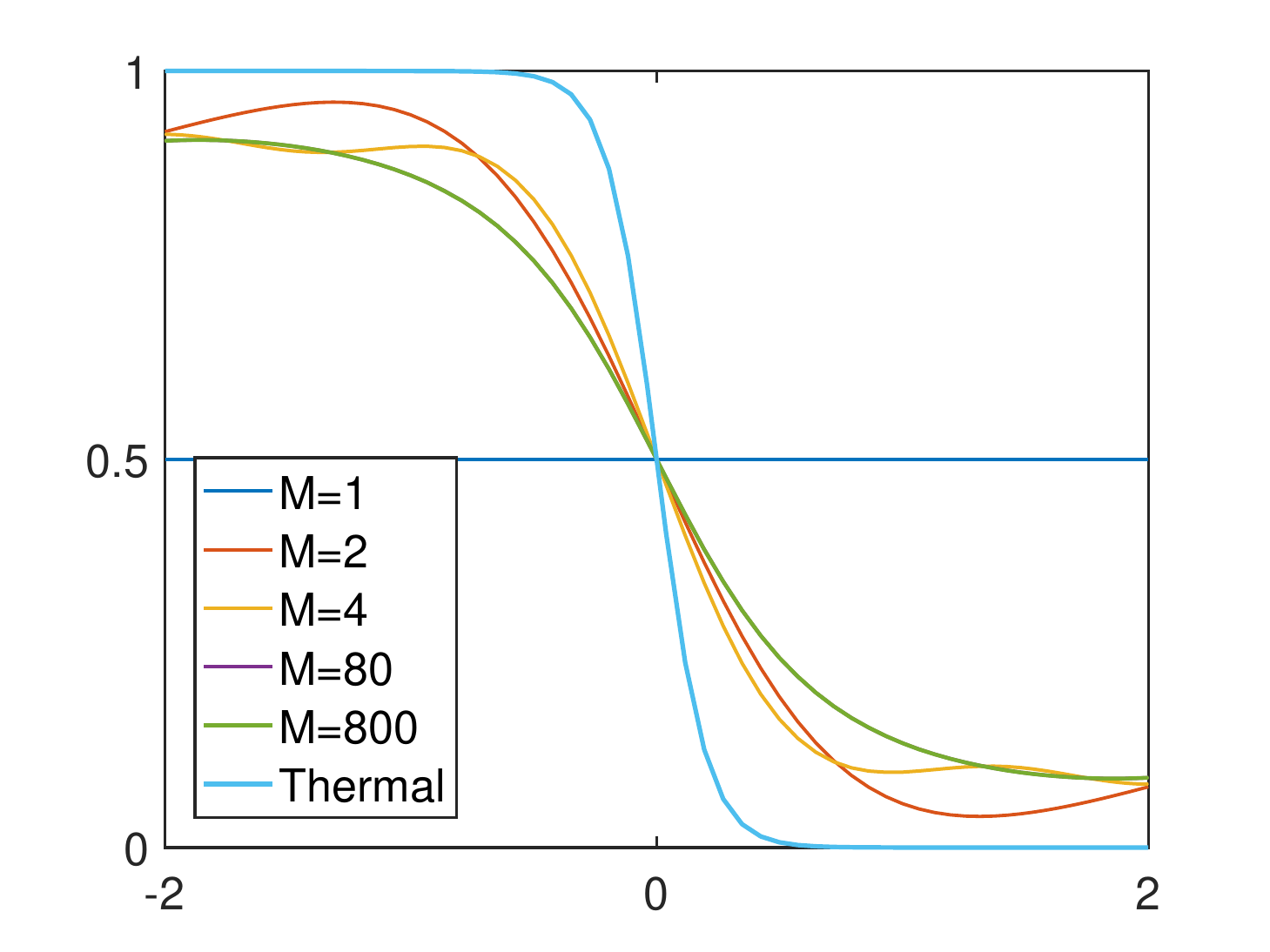}
\put(0,120){\large $g_n$}
\put(160,0){\large $\epsilon$}
\end{overpic}
\caption{Steady-state occupations of the single particle energy eigenstates, for
different lead sizes, as a function of the single particle state energies. Here: $t^{\prime}=1$, $\beta=10$, $\mu=0$, $\gamma=1$, $t_{L}=1.2$
and $N=80$. The thermalization error (Eq. (\ref{eq:therm_error}))
saturates to a value of $\Delta=0.14$ for $M>50$. }
\label{fig:therm_error_vs._system_size}
\end{figure}

In Fig.~\ref{fig:therm_error_vs._system_size} we show the occupation $g_n$ of one-particle system eigenstates as a function of energy, for a system with $N=80$ sites.  We take $t^{\prime}=1$, $t_L=1.2$, $\beta=10$, and constant rates $\gamma_{m}\equiv\gamma=1$.  The results are displayed for different values of the lead size $M$.

As expected from the previous section, for $M=1$ the occupation of states is independent of their energy.  This ceases to be true for $M>1$.  However, even for leads that are much larger than the system size,  $M\gg N$, the one-particle occupations $g_n$ are far from a thermal distribution with inverse temperature $\beta$. To quantify the error, we introduce the measure
\begin{equation}
\Delta\equiv\sqrt{\frac{1}{N}\sum_{n}\left(g_n-f_{{\rm FD}}\left(\epsilon_{n}^{S}\right)\right)^{2}}\label{eq:therm_error}
\end{equation}
where $\epsilon_{n}^{S}$ are the system one-particle energies,
see Eq. (\ref{eq:hamiltonian_system_eigenmodes}).   For the parameters considered above, the error already saturates to the large lead value for $M>50$ at $\Delta=0.14$.  This implies that the average deviation from the thermal occupation is $14\%$, even for an infinite lead.

\begin{figure}
\centering
\begin{overpic}[width = 0.45\textwidth]{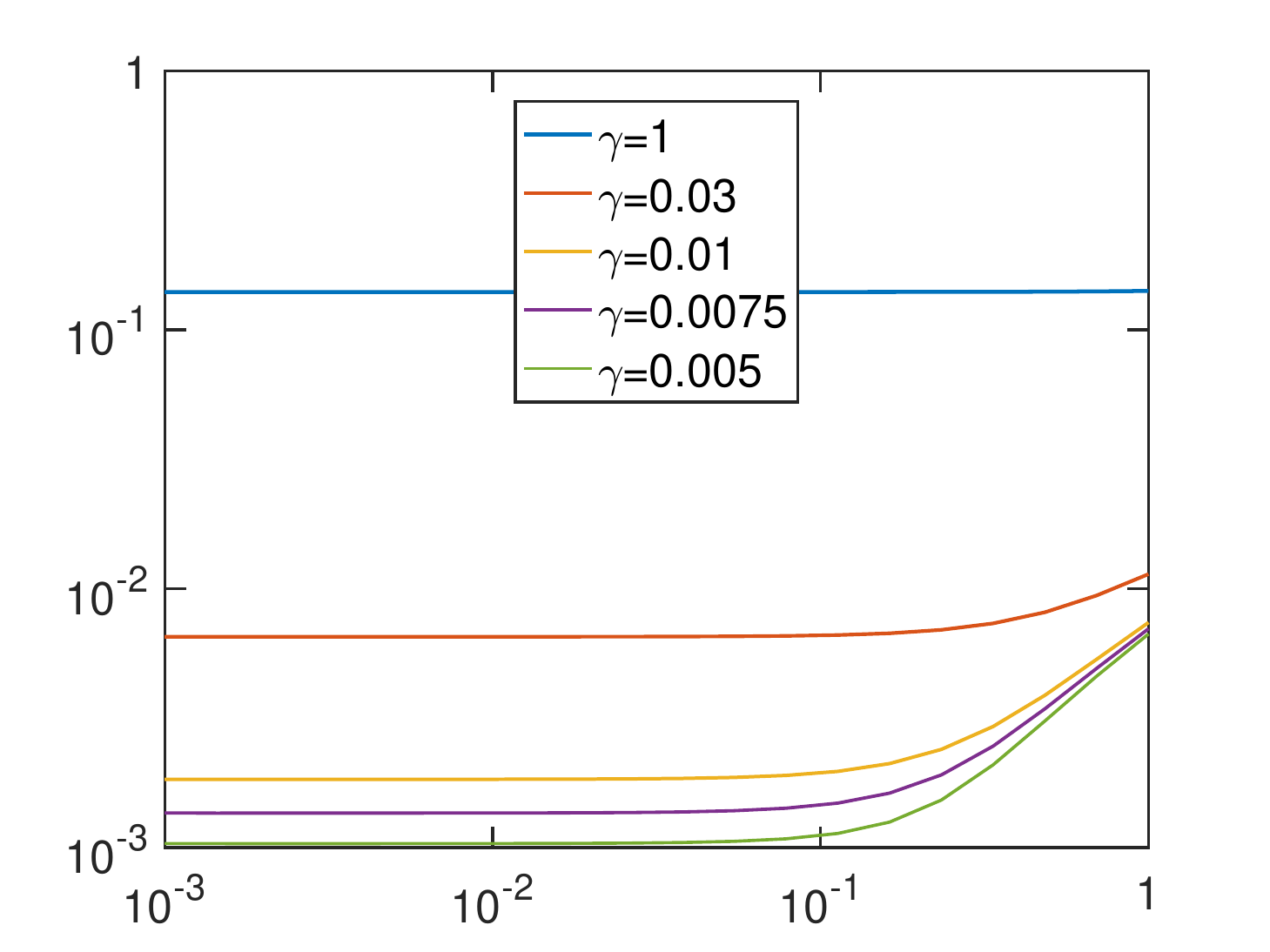}
\put(0,85){\large $\Delta$}
\put(110,0){\large $t'$}
\end{overpic}
\protect\caption{Thermalization error as a function of $t^{\prime}$, for different values of $\gamma$.
Here: $\beta=10$, $t_{L}=1.2$, $N=80$ and $M=400$.
\label{fig:therm_error_vs._tprime}}
\end{figure}

We now turn to investigate the effect of varying the values of the dissipation rate $\gamma$ and the system-lead coupling $t'$.  Figure  \ref{fig:therm_error_vs._tprime} shows the error $\Delta$ as a function of $t'$, for different values of $\gamma$, and a large lead.  One can see that $\Delta$ increases monotonically with $t'$, as one may expect from general statistical physics arguments: large values of $t'$ increase the subextensive corrections to the system's energy.  Note, however, that the error saturates at small $t'$ to a $\gamma$-dependent value, which is monotonically increasing with $\gamma$.  This implies that, even for leads that are much larger than the system size, the only hope of achieving thermalization is by taking both $t'$ and $\gamma$ to be small.

At first glance, the requirement of small $\gamma$ is counterintuitive, as one may have expected a strong dissipative coupling to lead to better equilibration.  However, as we show in the next section, a large $\gamma$ introduces an uncertainty in the system's energy, which reduces its ability to equilibrate at the desired temperature.  

As discussed at the end of Sec.~\ref{sec:model}, in microscopic derivations of the dissipative term, it is natural to consider the limit $t'\ll\gamma$.  Therefore, in what follows we will take the limit $t'\to 0$ first.  In this limit, we derive closed form expressions for the steady state density matrix and study the dependence of $\Delta$ on $\gamma$ and other parameters.

\section{Weak Lead Coupling  ($t'\to 0$)}

\subsection{Degenerate Perturbation Theory -- Integrable Case}
\label{sec:weakCoupling}

In this section, we will show that in the limit of infinitesimal coupling between system and lead, $t'\to 0$, the steady-state density matrix is given by
\begin{eqnarray}
\rho & \equiv & \prod_{m}\left(f_{m}b_{m}^{\dagger}b_{m}+\left(1-f_{m}\right)b_{m}b_{m}^{\dagger}\right)\nonumber \\
 &  & \times\prod_{n}\left(g_{n}c_{n}^{\dagger}c_{n}+\left(1-g_{n}\right)c_{n}c_{n}^{\dagger}\right)\label{eq:rhotp0}
\end{eqnarray}
where $0\leq g_{n}\leq1$ are the occupations of the system eigenstates, given by
\begin{equation}
g_n=\frac{\sum_{m}\left|t_{mn}^{\prime}\right|^{2}Q_{nm} f_{m}}{\sum_{m}\left|t_{mn}^{\prime}\right|^{2}Q_{nm}}\label{eq:occupations_perturbation_theory}
\end{equation}
where $Q_{nm}=\Upsilon\!\left(\epsilon_{m}^{L}-\epsilon_{n}^{S}\right)$ and
\begin{equation}
\Upsilon(\Delta\epsilon)=\frac{\gamma}{\left(\Delta\epsilon\right)^{2}+\gamma^{2}/4}.\label{eq:lorentzian}
\end{equation}
Here we assume, for simplicity, that $\gamma_m=\gamma$.
The hopping amplitudes $t_{mn}^{\prime}$ are defined through
\begin{equation}
H_{\rm int}=-\sum_{mn}t_{mn}^{\prime}b_{m}^{\dagger}c_{n}+{\rm h.c.},
\end{equation}
that is, $t'_{mn}$ is the coupling between lead and system in the energy basis of both.  Note that the density matrix in Eq. (\ref{eq:rhotp0}) is diagonal  in the energy basis, as must be the case in order for $\rho$ to commute with the system Hamiltonian in the steady state.

Equation (\ref{eq:occupations_perturbation_theory}) can be interpreted
as a balance equation between particles hopping in and out of the system state $n$, via particle exchange with the lead:
\begin{equation}
0=\left(1-g_{n}\right)R_{n}^{{\rm in}}-g_{n}R_{n}^{{\rm out}}\label{eq:balanceSL}
\end{equation}
where the rates $R_{n}^{{\rm in}}$, $R_{n}^{{\rm out}}$ include
all possible hopping processes:
\begin{align}
R_{n}^{{\rm in}} & =\sum_{m}|t'_{nm}|^{2}Q_{nm}\,f_{m}\nonumber \\
R_{n}^{{\rm out}} & =\sum_{m}|t'_{nm}|^{2}Q_{nm}\,(1-f_{m})\label{eq:ratesSL}
\end{align}
The factors $1-g_{n}$ (in Eq. (\ref{eq:balanceSL})) and $1-f_{m}$
(in Eq. (\ref{eq:ratesSL})) appear due to Pauli exclusion. 
In the limit $\gamma\to0$, $Q_{nm}\to2\pi\delta\left(\epsilon_{n}^{S}-\epsilon_{m}^{L}\right)$,
and these rates reduce to the standard result expected from Fermi's
Golden Rule. More generally, for non-zero $\gamma_{m}$, the dissipative
term broadens the energy-conserving $\delta$-function into a Lorentzian
$Q_{nm}$.

This imposes constraints on when the equilibrium density matrix,
\begin{eqnarray}
\rho_S = \prod_{n}\left(f_{\rm FD} \left(\epsilon_n^S\right)c_{n}^{\dagger}c_{n}+\left(1-f_{\rm FD} \left(\epsilon_n^S\right)\right)c_{n}c_{n}^{\dagger}\right),
\end{eqnarray}
(corresponding to taking $g_n=f_{\rm FD}(\epsilon_n^S)$ in Eq. (\ref{eq:rhotp0})) can be obtained.  Specifically, in the limit $\gamma\to 0$, we require that the lead is infinite and its bandwidth is larger than or equal to that of the system.  To understand these conditions, note that when $\gamma\to 0$, the Lorentzian $Q_{mn}$ selects the energy $\epsilon_n^S=\epsilon_m^L$.  When the lead is infinite and its bandwidth is large, one is guaranteed to find a state in the lead that matches each level in the system.

Equation (\ref{eq:occupations_perturbation_theory}) is much simpler to use than the third quantization method, as it only requires a diagonalization of the system Hamiltonian.  It applies to the class of integrable systems described above, and to more general tight-binding models, and only relies on the weak coupling between system and lead.  It can be generalized to systems with anomalous fermion hopping $c_j^\dagger c_{j'}^\dagger$, which are relevant for superconducting systems and to quantum Ising chains.

The full derivation of the above results is given in Appendix \ref{appen:PertTheory}. Here, we outline its main steps. This derivation is similar to previous studies of the perturbative approach to the Lindblad equation (see Refs. \onlinecite{Lenarcic2017, Guimaraes2016, Benatti2011, Li2013, Li2016}), with appropriate modifications relevant to our physical setup.
First, we write the Lindblad equation for the steady state in the form:
\begin{eqnarray}
0 & = & \hat{H}_{0}\rho+\hat{\Gamma}\rho+\hat{V}\rho\label{eq:lindblad_steady_state}
\end{eqnarray}
with the superoperators $\hat{H}_{0}$ and $\hat{V}$ defined by,
\begin{eqnarray}
\hat{H}_{0}\rho & = & -i\left[H_{S}+H_{L},\rho\right],\nonumber \\
\hat{V}\rho & = & -i\left[H_{{\rm int}},\rho\right],
\end{eqnarray}
and $\hat{\Gamma}$ given in Eq.~(\ref{eq:lindblad_thermalization}).  We are interested in the steady state $\rho$ in the limit of weak system-lead coupling $\hat{V}\to 0$.

Let's consider first the case $\hat{V}=0$.  Then, the density matrix $\rho$ factorizes into system and lead components.  The lead component is thermal by construction.  The system component is a steady state, provided it commutes with $H_S$.  This implies that $\rho$ is of the form in Eq. (\ref{eq:rhotp0}), where the parameters $g_n$ are at this point arbitrary.  For simplicity, we assume $H_{S}$ has a non-degenerate spectrum.

In order to obtain the density matrix to leading order in $t'$, the coefficients $g_n$ have to be determined using degenerate perturbation theory, within the Hilbert space of states of the form in Eq. (\ref{eq:rhotp0}).  The details are given in Appendix \ref{appen:PertTheory}.

\subsection{Fluctuation-Dissipation Relation}

It has been shown that the Lindblad equation does not in general satisfy fluctuation-dissipation relations \cite{Talkner1986,Ford1996,Keeling}.  This is true even in cases where the Lindblad operator acts globally on the whole system and when the steady state is Gibbs. Fluctuation-dissipation relations hold only in the weak coupling limit and are violated otherwise. This can be understood by recalling that unitary dynamics together with Gibbs initial condition for the density matrix ensure the fluctuation-dissipation relations. The dynamics of the system are unitary over time scales shorter than $1/\gamma$, and therefore the fluctuation-dissipation relations become exact in the weak coupling limit $\gamma\to 0$.

In our settings we saw that a small $\gamma$ (and $t'$) limit is required for obtaining a Gibb density matrix. With the above in mind this automatically ensures the existence of fluctuation-dissipation relations.  

\subsection{Error Analysis}
\label{sec:error}

We have shown that in order to obtain thermalization, we must take the limit of small $t'$ and $\gamma$, while holding a large lead $M\gg N$.  It is interesting to quantify the error.  As argued above, it is natural to consider the limit $t'\ll \gamma$.  Therefore, we estimate the error as a function of $\gamma$, to leading order in $t'\to 0$, using the results of Sec.~\ref{sec:weakCoupling}.  Throughout, we work with an infinite lead of infinite bandwidth.

Using Eq. (\ref{eq:occupations_perturbation_theory}), we obtain
\begin{equation}
g\left(\epsilon_{n}^{S}\right)=\frac{\int d\epsilon\, \Upsilon\!\left(\epsilon-\epsilon_{n}^{S}\right) f_{\rm FD}\left(\epsilon\right)}{\int d\epsilon\, \Upsilon\!\left(\epsilon-\epsilon_{n}^{S}\right)}.\label{eq:occupations_perturbation_theory_inflead}
\end{equation}
Here, as above, we assume that the system couples to the lead on a single site, and that the lead has periodic boundary conditions.  Then, $|t'_{mn}|^2$ is independent of $m$, and it drops out of the sum in Eq. (\ref{eq:occupations_perturbation_theory}).  For simplicity, we have also assumed that the density of states varies slowly relative to both $\gamma$ and $\beta^{-1}$, which are also taken to be small relative to the system bandwidth.  Then, the density of states also drops out, and the integrals above can be taken to extend from $-\infty$ to $\infty$.  In this case, the denominator becomes $\int d\epsilon\, \Upsilon\!\left(\epsilon-\epsilon_{n}^{S}\right)=2\pi$. 

\begin{figure}
\centering
\begin{overpic}[width = 0.45\textwidth]{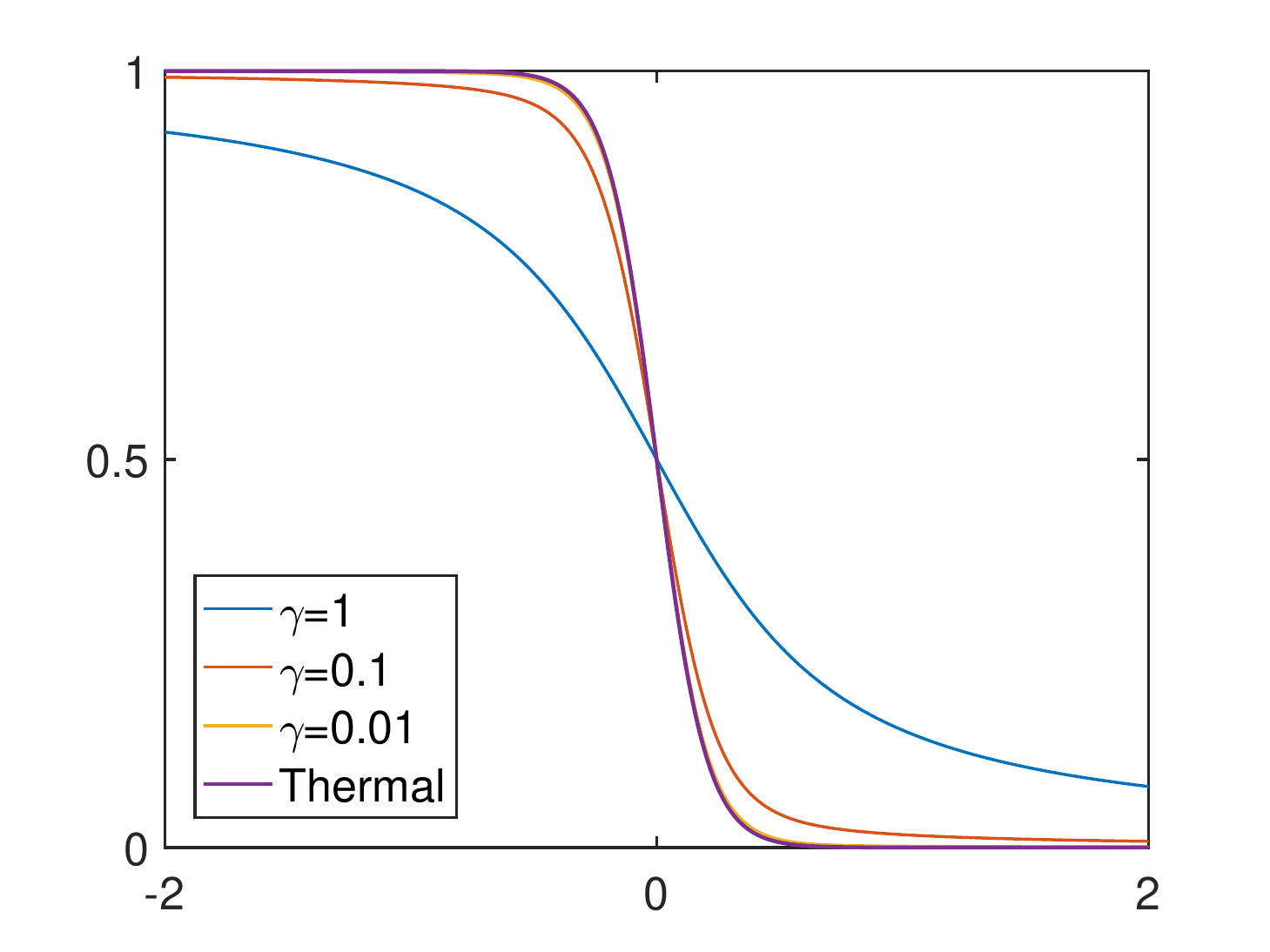}
\put(0,120){\large $g_n$}
\put(160,0){\large $\epsilon$}
\end{overpic}
\protect\caption{Steady-state occupations of the single particle energy eigenstates for
$\beta=10$, in the limit $t^{\prime}\rightarrow0$ and 
an infinite size lead. \label{fig:therm_error_tprimezero_beta10}}
\end{figure}

Equation (\ref{eq:occupations_perturbation_theory_inflead}) shows that the occupation is smeared by the Lorentzian resulting from the dissipative terms.  This implies a condition for thermalization, $\gamma\ll \beta^{-1}$.  Figure \ref{fig:therm_error_tprimezero_beta10} shows the occupation function for different values of $\gamma$, at a fixed value $\beta=10$.  As can be clearly seen, only for $\gamma\beta\ll 1$ does the occupation approximate well the thermal distribution.

\begin{figure}
\centering
\begin{overpic}[width = 0.45\textwidth]{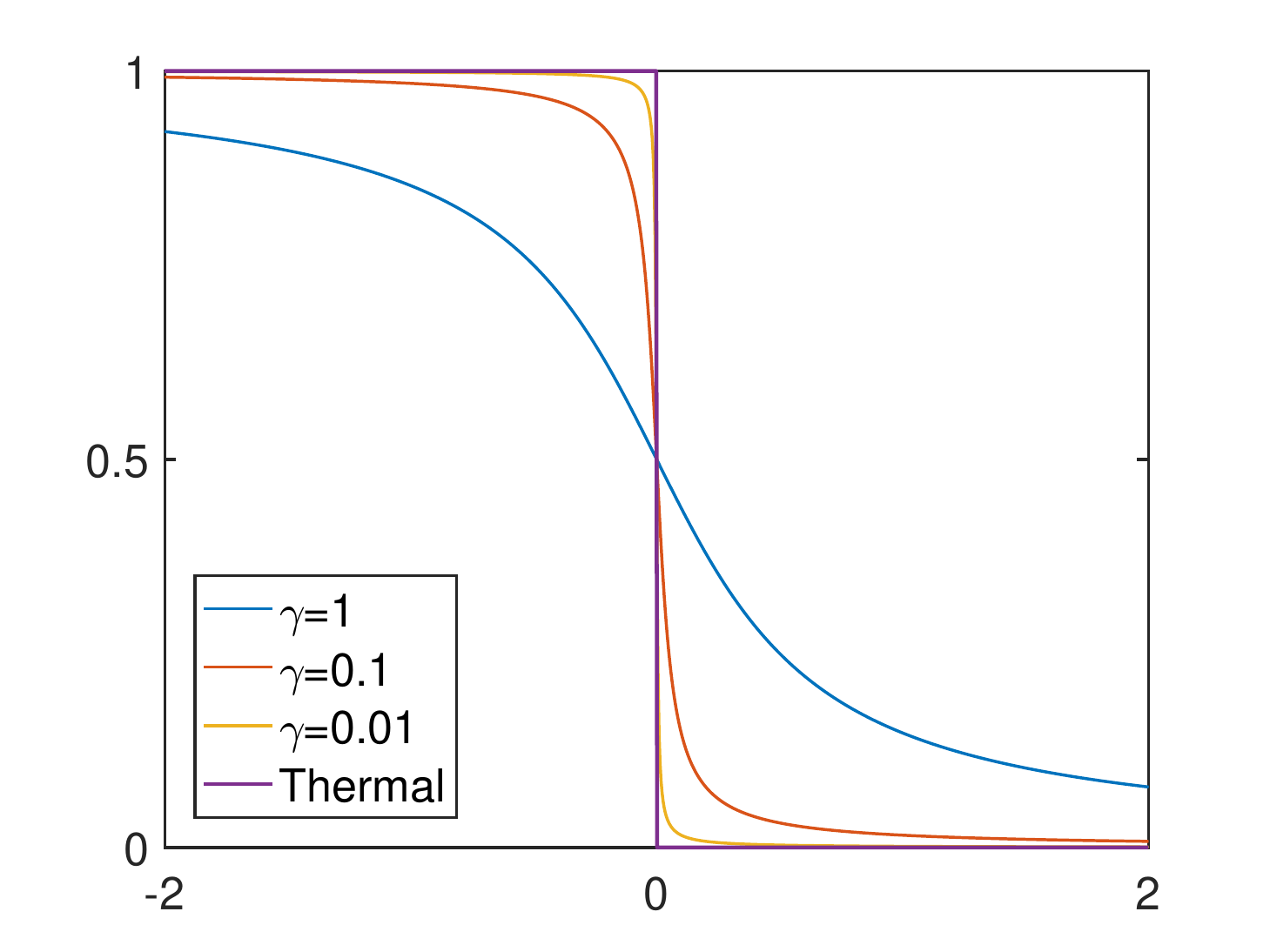}
\end{overpic}
\protect\caption{Steady-state occupations of the single particle energy eigenstates at
zero temperature, in the limit $t^{\prime}\rightarrow0$ and for
an infinite lead size. \label{fig:therm_error_tprimezero_betainf}}
\end{figure}

\begin{figure}
\centering
\begin{overpic}[width = 0.45\textwidth]{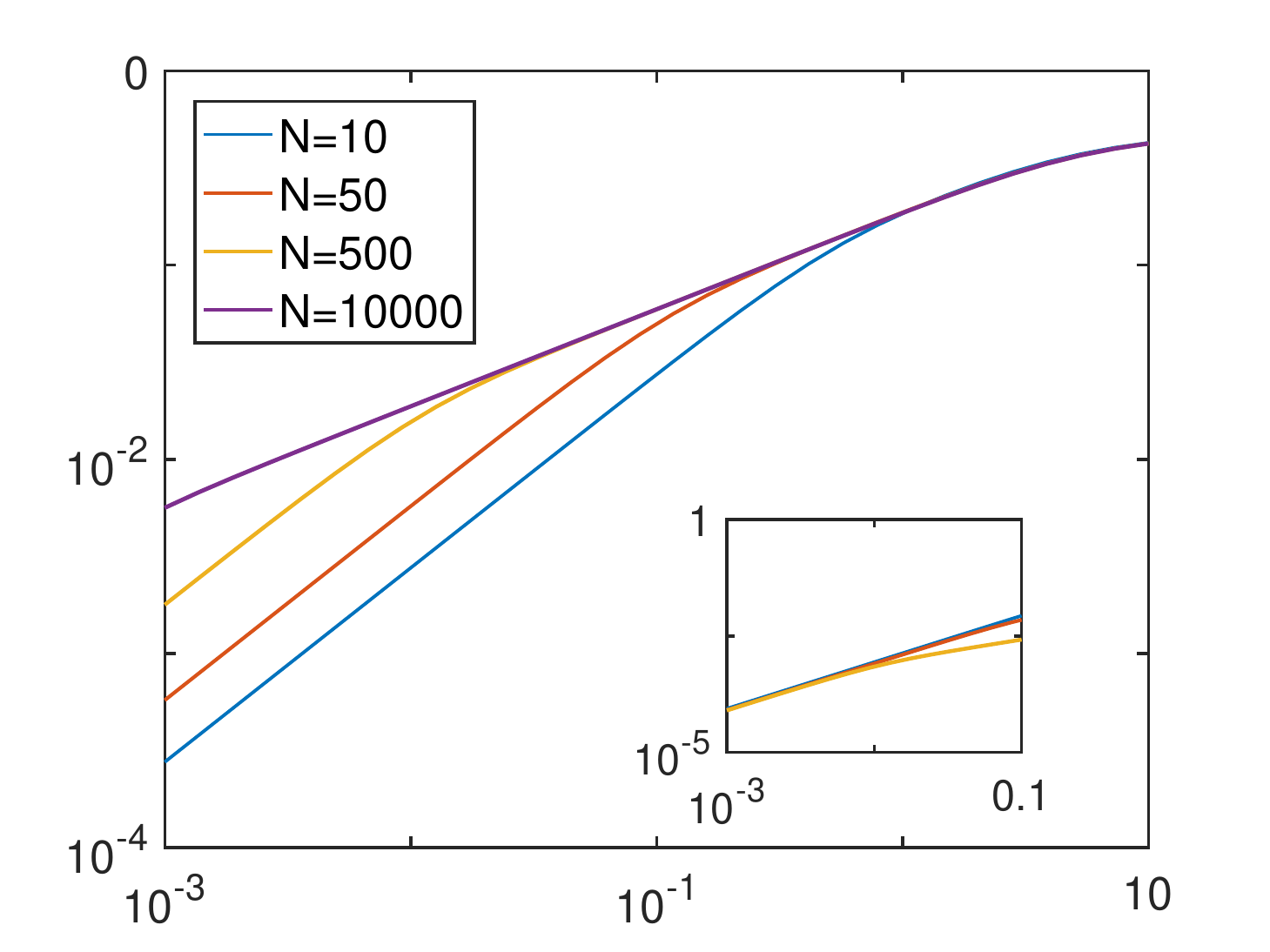}
\put(0,120){\large $\Delta$}
\put(115,60){$\frac{\Delta}{\sqrt{N}}$}
\put(160,0){\large $\gamma$}
\end{overpic}

\protect\caption{Thermalization error at $\beta=\infty$ as a function of $\gamma$ for different
values of $N$, in the limit $t^{\prime}\rightarrow0$ and for
an infinite lead size. Inset: the same data, scaled with $\frac{1}{\sqrt{N}}$. Note that the plot for $N=10000$ has been omitted from the inset, since it is practically an infinite system for the values of $\gamma$ considered here. \label{fig:therm_error_tprimezero_betainfinite}}
\end{figure}

In the opposite limit, when $\beta\gamma\gg1$, Eq.~(\ref{eq:occupations_perturbation_theory_inflead})
evaluates to 
\begin{equation}
g\left(\epsilon_{n}^{S}\right)=\frac{1}{2}-\frac{1}{\pi}\tan^{-1}\left(\frac{2\epsilon_{n}^{S}}{\gamma}\right).\label{eq:occupations_perturbation_theory_inflead_infbeta}
\end{equation}
Note that this result is independent of $\beta$, as one may expect
in this regime. The occupation is controlled by the Lorentzian, which
has broad tails, as shown in Fig. \ref{fig:therm_error_tprimezero_betainf}. This leads to a slow decay of the occupation, which
behaves as $\sim\frac{\gamma}{2\pi\epsilon_{n}^{S}}$ well above the
chemical potential of the lead, and therefore cannot be described
by a Fermi distribution with an effective temperature. The thermalization
error in this regime is dictated by the system size. As clearly seen
in Fig. \ref{fig:therm_error_tprimezero_betainfinite}, in the limit
$\gamma N\ll1$ it behaves as $\Delta^{2}\propto\gamma^{2}$, whereas
in the opposite limit $\gamma N\gg1$ it behaves as $\Delta^{2}\propto\gamma$.
To explain this, we note that for finite systems, the energy level
spacing scales as $\frac{1}{N}$. Therefore, the main contributions
to the error near the Fermi surface ($\epsilon_{n}^{S}=0$) are simply
\begin{equation}
g\left(\epsilon_{n}^{S}\right)_{\gamma N\ll1}\sim\begin{cases}
1-\gamma N & \epsilon_{n}^{S}<0\\
\gamma N & \epsilon_{n}^{S}>0
\end{cases}
\end{equation}
and clearly $\Delta^{2}\propto\gamma^{2}N$ (see Fig. \ref{fig:therm_error_tprimezero_betainfinite} inset). On the other hand, for
infinite systems the summation can be replaced by an integral,
\begin{equation}
\Delta^{2}=2\cdot\int_{0}^{W}d\epsilon D\left(\epsilon\right)\left(\frac{1}{2}-\frac{1}{\pi}\tan^{-1}\left(\frac{2\epsilon}{\gamma}\right)\right)^{2}
\end{equation}
with $D\left(\epsilon\right)$ being the density of states and $W$ is
a characteristic bandwidth, of order $1$. Assuming that the density
of states does not change abruptly, namely $D^{\prime}\left(\epsilon\right)\ll\frac{1}{\gamma}$,
it can be approximated to a constant. Then, doing a simple change
of variables and noting that $\frac{W}{\gamma}\gg1$ leads to
$\Delta^{2}\propto\gamma$.

\begin{figure}
\centering
\begin{overpic}[width = 0.45\textwidth]{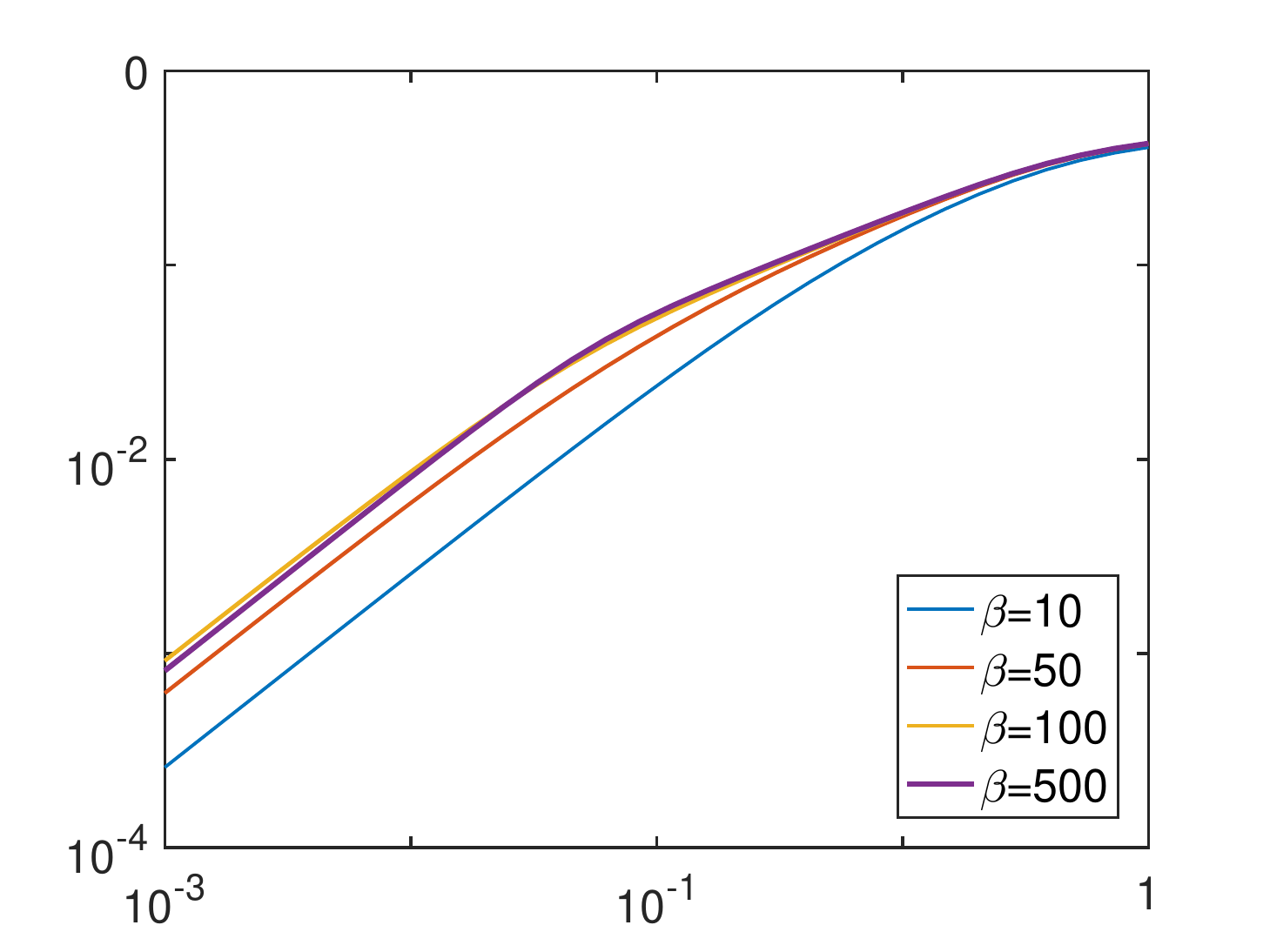}
\put(0,120){\large $\Delta$}
\put(160,0){\large $\gamma$}
\end{overpic}
\protect\caption{Thermalization error as a function of $\gamma$ for different values of $\beta$,
in the limit $t^{\prime}\rightarrow0$ and an infinite size lead. System size is finite and chosen as $N=100$.\label{fig:therm_error_tprimezero_betafinite}}
\end{figure}

In Fig. \ref{fig:therm_error_tprimezero_betafinite}, we plot the error for finite temperatures, but still
low enough such that $\beta\gg1$.
The error saturates for large values of $\beta$, as well as for large
values of $\gamma$.

\section{Generalization to non-Integrable Systems}

In the non-integrable case, in order to study many-body systems, we write the system Hamiltonian in terms
of the many-body eigenstates:
\begin{equation}
H_{S}=\sum_{\nu=1}^{2^{N}}E_{\nu}^{S}\ket{\psi_{\nu}}\bra{\psi_{\nu}}
\end{equation}
with $\ket{\psi_{\nu}}$ a many-body eigenstate and $E_{\nu}^{S}$ its
corresponding energy. For simplicity, we keep studying particle
number conserving Hamiltonians, such as Eq. (\ref{eq:HSnon_int}). In this case, each eigenstate has
a well defined particle number, such that 
\begin{equation}
\hat{\mathcal{N}}\ket{\psi_{\nu}}=\mathcal{N}_{\nu}\ket{\psi_{\nu}}
\end{equation}
with $\mathcal{\hat{N}}=\sum_{i=1}^{N}c_{i}^{\dagger}c_{i}$ and $\mathcal{N}_{\nu}$
an integer between $0$ and $N$. 

The coupling Hamitlonian, Eq. (\ref{eq:hamiltonian_coulping}), exchanges
one particle between the system and the lead, and therefore can be
written in terms of energy eigenstates as follows:
\begin{equation}
H_{{\rm int}}=\sum_{m,\nu,\nu^{\prime}}T_{m\nu\nu^{\prime}}b_{m}^{\dagger}\ket{\psi_{\nu^{\prime}}}\bra{\psi_{\nu}}+{\rm h.c.}
\end{equation}
with 
\begin{equation}
T_{m\nu\nu'}=-\frac{e^{-2\pi imM}}{\sqrt{M}} t' \bra{\psi_{\nu^{\prime}}}c_{1}\ket{\psi_{\nu}} \;.
\end{equation}
This matrix element
connects eigenstates which differ by one particle, namely $\mathcal{N}_{\nu}=1+\mathcal{N}_{\nu^{\prime}}$.

As before, we start by studying the case of $H_{{\rm int}}=0$. Then, the steady state density matrix factorizes into a system component and a thermal density matrix for the lead. Assuming that there are no degeneracies, it can therefore be written as follows:
\begin{eqnarray}
\rho & = & \prod_{m=1}^{M}\left(f_{m}b_{m}^{\dagger}b_{m}+\left(1-f_{m}\right)b_{m}b_{m}^{\dagger}\right)\nonumber \\ &  & 
\otimes\sum_{\nu=1}^{2^N}G_{\nu}\ket{\psi_{\nu}}\bra{\psi_{\nu}}\label{eq:steady_state_non_integrable}
\end{eqnarray}
with $0\le G_{\nu}\le 1$ arbitrary and subject to the normalization
condition $\sum_{\nu}G_{\nu}=1$.   As a side remark, we point
out that degeneracies can arise due to the particle-hole symmetry in
our system at $\mu=0$. These can be simply taken into account by a simultaneous diagonalization
of $H_{S}$ and $\hat{\mathcal{\mathcal{N}}}$ \cite{Cardoso1996}.

To proceed we carry out a degenerate perturbation theory in $t'$. This fixes the values of $G_{\nu}$. The analysis is
very similar to the integrable case, as presented in Appendix \ref{appen:PertTheory},
and results in the following equation for each state $\ket{\psi_{\nu}}$:
\begin{eqnarray}
0 & = & \sum_{m,\nu^{\prime}}\left|T_{m\nu\nu^{\prime}}\right|^{2}Q_{m\nu\nu^{\prime}}\left(f_{m}G_{\nu^{\prime}}-\left(1-f_{m}\right)G_{\nu}\right)+\nonumber \\
 & \, & \sum_{m,\nu^{\prime\prime}}\left|T_{m\nu^{\prime\prime}\nu}\right|^{2}Q_{m\nu^{\prime\prime}\nu}\left(\left(1-f_{m}\right)G_{\nu^{\prime\prime}}-f_{m}G_{\nu}\right)\label{eq:secular_non_integrable}
\end{eqnarray}
where the summation runs over eigenstates such that 
\begin{align}
\mathcal{N}_{\nu^{\prime}} & =\mathcal{N}_{\nu}-1\nonumber \\
\mathcal{N}_{\nu^{\prime\prime}} & =\mathcal{N}_{\nu}+1
\end{align}
In addition, 
\begin{eqnarray}
Q_{m\nu\nu^{\prime}}=\varUpsilon\left(\epsilon_{m}^{L}-\left(E_{\nu}^{S}-E_{\nu^{\prime}}^{S}\right)\right)
\end{eqnarray}
with $\varUpsilon\left(\Delta\epsilon\right)$ defined in Eq. (\ref{eq:lorentzian}).
The steady state values of $G_{\nu}$ are then obtained by solving the
$2^{N}$ coupled linear equations of Eq. (\ref{eq:secular_non_integrable}).

Equation (\ref{eq:secular_non_integrable}) genealizes Eq. (\ref{eq:balanceSL}) to a many-body context. It can be rephrased
as a global balance equation for the flow in and out of the state
$\ket{\psi_{\nu}}$:
\begin{align}
0 & =\sum_{\nu^{\prime}}\left(G_{\nu^{\prime}}R_{\nu^{\prime}\rightarrow\nu}
-G_{\nu}R_{\nu\rightarrow\nu^{\prime}}
\right)\nonumber \\
 & +\sum_{\nu^{\prime\prime}}\left(G_{\nu^{\prime\prime}}R_{\nu^{\prime\prime}\rightarrow\nu}
-G_{\nu}R_{\nu\rightarrow\nu^{\prime\prime}}
\right)\label{eq:balance_nonint}
\end{align}
with $G_{\nu}$ again being the probability for the eigenstate $\nu$,
and where the rates $R$ sum over all possible particle exchanges with the lead:
\begin{align}
R_{\nu^{\prime}\rightarrow\nu} & =\sum_{m}\left|T_{m\nu\nu^{\prime}}\right|^{2}Q_{m\nu\nu^{\prime}}f_{m}\nonumber \\
R_{\nu\rightarrow\nu^{\prime}} & =\sum_{m}\left|T_{m\nu\nu^{\prime}}\right|^{2}Q_{m\nu\nu^{\prime}}\left(1-f_{m}\right)\nonumber \\
R_{\nu^{\prime\prime}\rightarrow\nu} & =\sum_{m}\left|T_{m\nu^{\prime\prime}\nu}\right|^{2}Q_{m\nu^{\prime\prime}\nu}\left(1-f_{m}\right)\nonumber \\
R_{\nu\rightarrow\nu^{\prime\prime}} & =\sum_{m}\left|T_{m\nu^{\prime\prime}\nu}\right|^{2}Q_{m\nu^{\prime\prime}\nu}f_{m \;.}\label{eq:balance_nonint_rates}
\end{align}
Note that the expression for the rates depends on whether a particle is being added or removed into the lead. Since Eq. (\ref{eq:balance_nonint}) is a balance equation for a probability conserving process, there is always a solution for the probabilities $G_\nu$.  If the process is ergodic, then the solution is unique.

As in the integrable case, the system thermalizes for $\gamma\to0$.  In this limit, $Q_{m\nu\nu^{\prime}}\to2\pi\delta\left(\epsilon_{m}^{L}-\left(E_{\nu}^{S}-E_{\nu^{\prime}}^{S}\right)\right)$
and we obtain the many-body equivalent of Fermi's Golden Rule. Then, for an infinite lead where the bandwidth is infinite, the rates in Eq. (\ref{eq:balance_nonint_rates}) satisfy detailed balance:
\begin{equation}
\frac{R_{\nu'\to\nu}}{R_{\nu\to\nu'}}=e^{-\beta\left(E_{\nu}^{S}-E_{\nu^{\prime}}^{S}-\mu\right)}\label{eq:detailed_balance_nonint}
\end{equation}
and similarly for the ratio $\frac{R_{\nu\to\nu''}}{R_{\nu''\to\nu}}$.  Hence, the steady state probabilities are \begin{equation}
G_\nu=G^{\rm thermal}_\nu\equiv \frac{1}{\mathcal{Q}} e^{-\left(\beta E_{\nu}^{S}-\mu {\mathcal{N}}_\nu\right)}
\end{equation}
where $\mathcal{Q}$ is the grand-canonical partition function: 
\begin{equation}
\mathcal{Q}\equiv\sum_{\nu}e^{-\left(\beta E_{\nu}^{S}-\mu {\mathcal{N}}_\nu\right)}
\end{equation}

\begin{figure}
\begin{centering}
\subfloat[\label{fig:therm_error_non_int_gamma_small}]{\begin{centering}
\begin{overpic}[width = 0.45\textwidth]{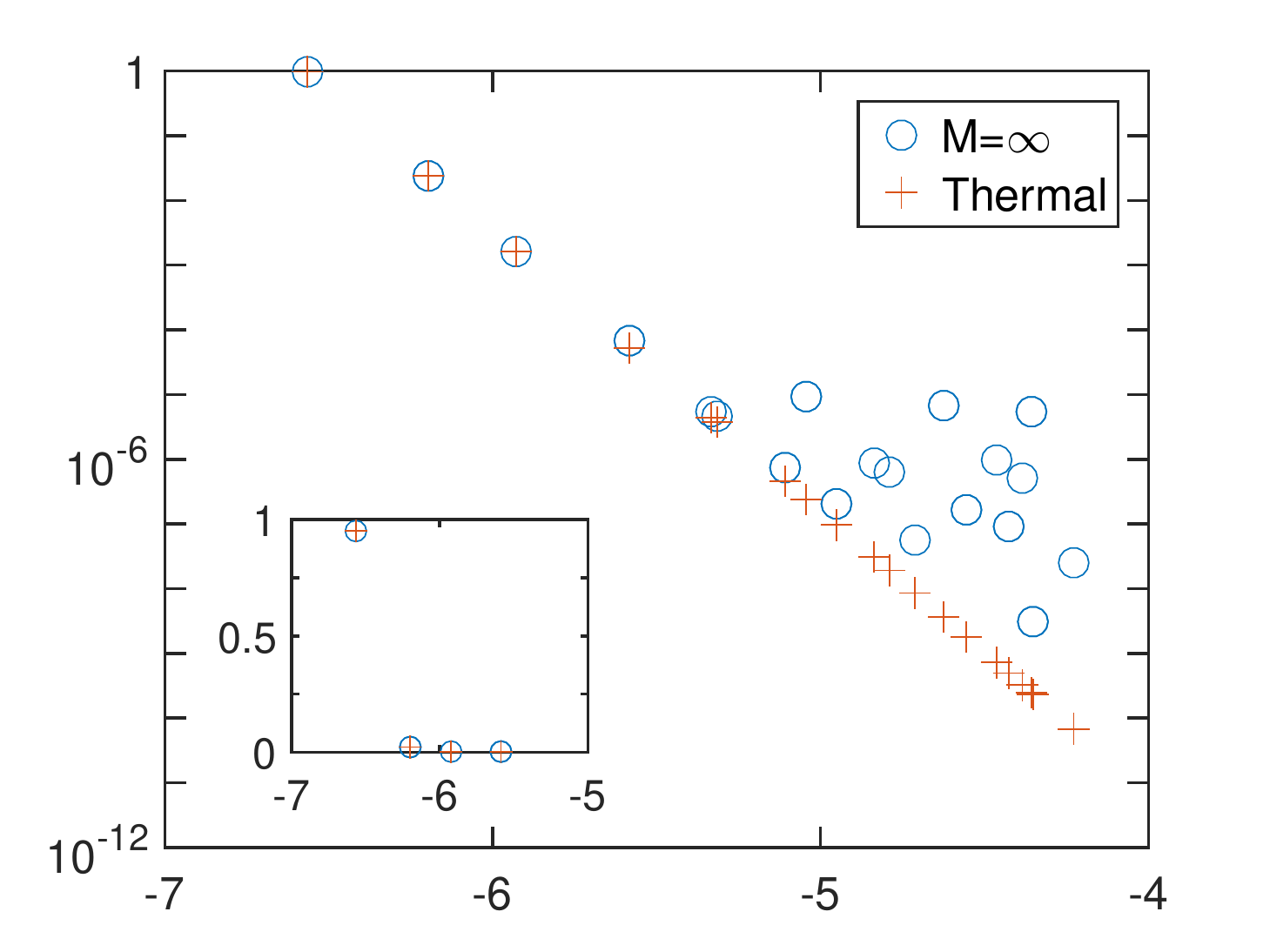}
\put(0,120){\large $G_{\nu}$}
\put(160,0){\large $E_{\nu}^{S}$}
\end{overpic}
\par\end{centering}

}\hfill{}\subfloat[\label{fig:therm_error_non_int_gamma_large}]{\begin{centering}
\begin{overpic}[width = 0.45\textwidth]{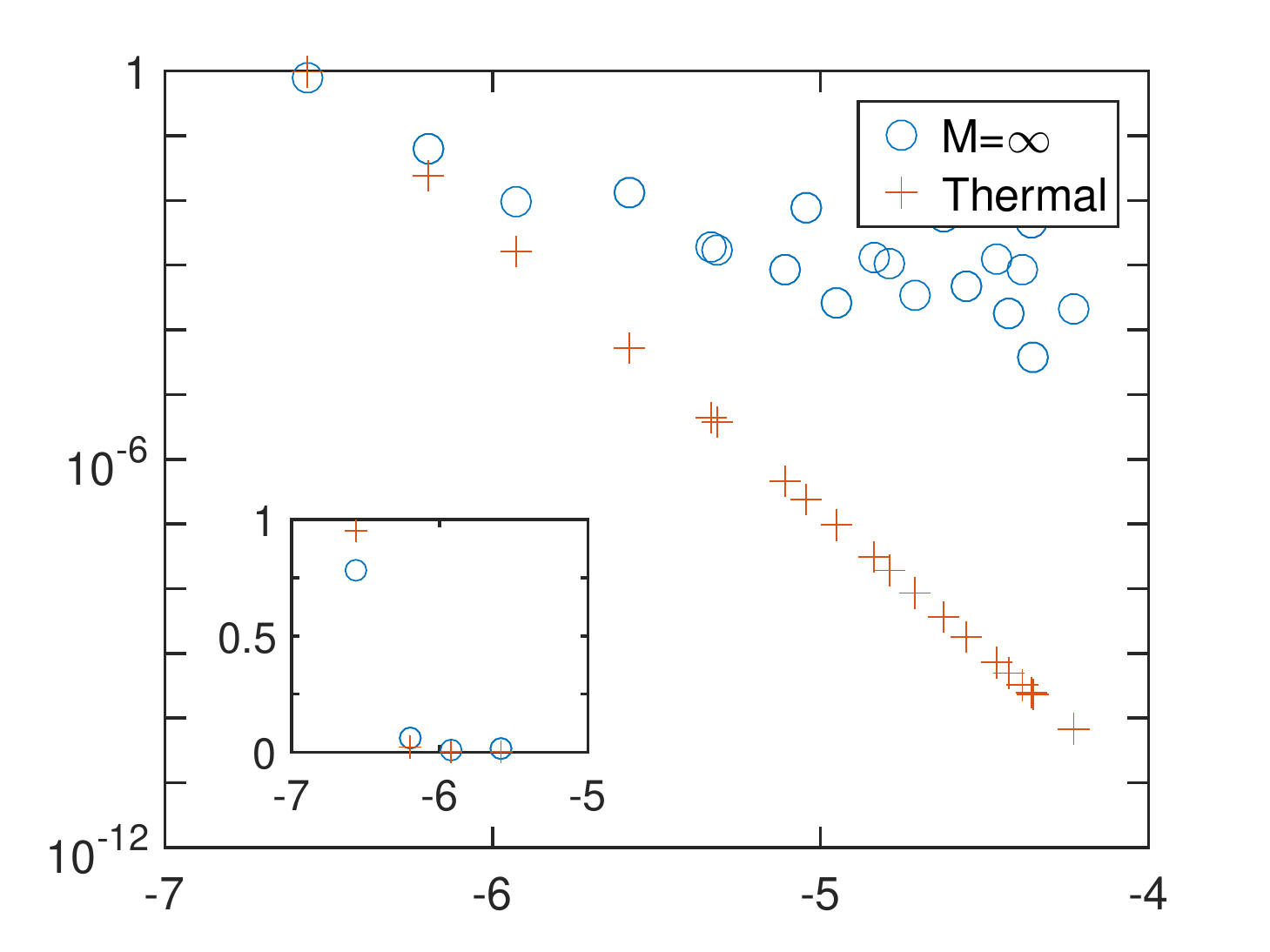}
\put(0,120){\large $G_{\nu}$}
\put(160,0){\large $E_{\nu}^{S}$}
\end{overpic}
\par\end{centering}
}
\par\end{centering}
\caption{Steady state occupations of the first $30$ many-particle energy eigenstates
in the limit $t^{\prime}\rightarrow0$ and for an infinite size lead. Here: $\beta=10$, $N=10$. Inset:
the same graph, on a linear scale, for the first $4$ eigenstates.
In (a), $\gamma=10^{-4}$, while in (b), $\gamma=0.1$.
\label{fig:therm_error_nonint}}
\end{figure}

Figure \ref{fig:therm_error_nonint} shows the steady state value of $G_\nu$, as a function of energy, for two different values of $\gamma$.  As expected, when $\gamma$ is small, the system is close to a thermal distribution.  Significant deviations from thermalization are only noticeable on a logarithmic scale, for excited states whose probability is $G_\nu\le 10^{-5}$.  On the other hand, when $\gamma$ is not small, there are significant deviations even for the ground state.

\begin{figure}
\begin{centering}
\begin{overpic}[width = 0.45\textwidth]{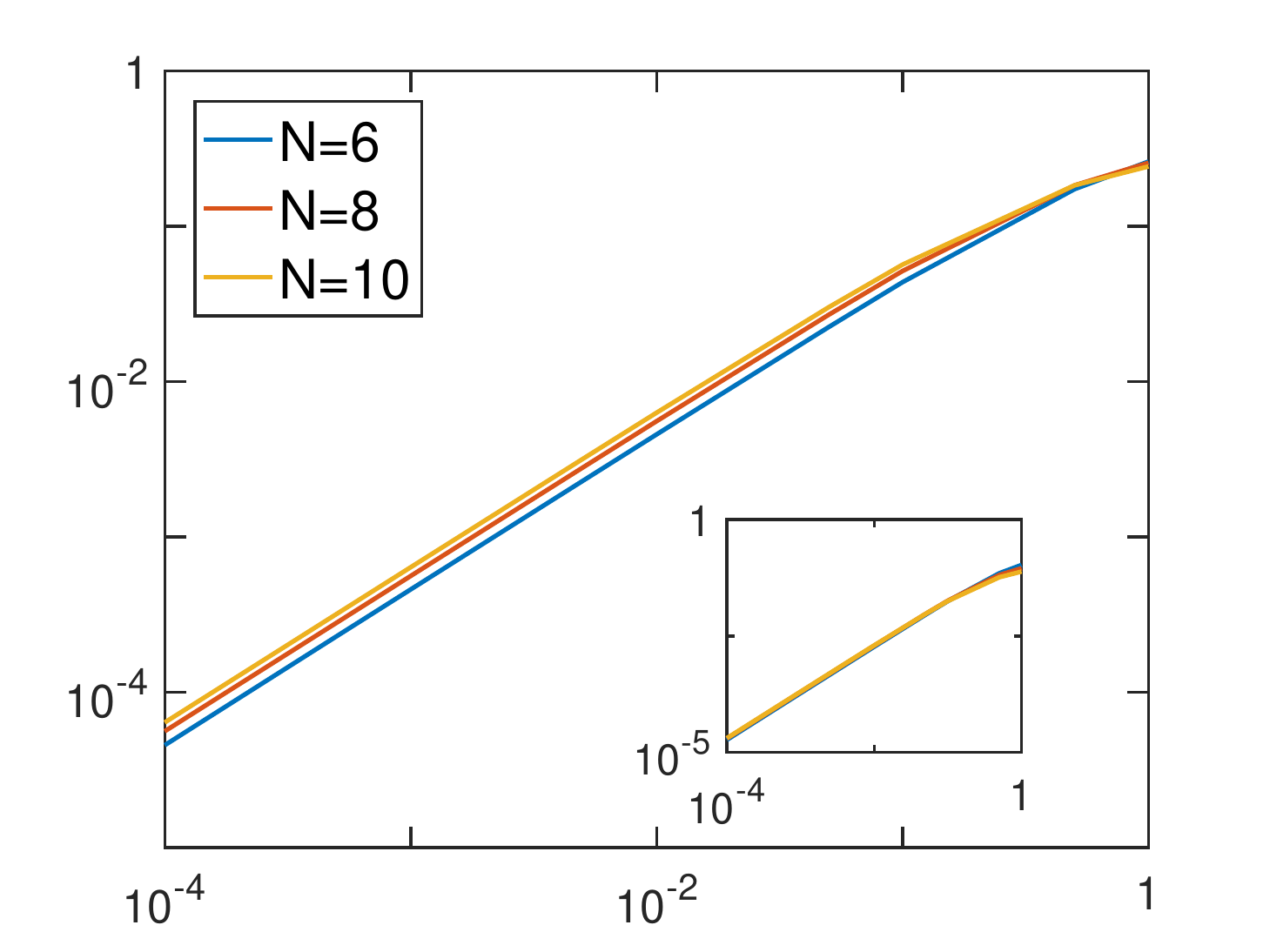}
\put(0,120){\large $\Delta$}
\put(115,60){$\frac{\Delta}{\sqrt{N}}$}
\put(160,0){\large $\gamma$}
\end{overpic}
\par\end{centering}
\caption{Many-body thermalization error (Eq. (\ref{eq:therm_error_nonint})) as a function of $\gamma$
for different system sizes and an infinite size lead (Here: $\beta=10$). Inset: the same data, scaled by $\frac{1}{\sqrt{N}}$. \label{fig:therm_error_non_int_vs._gamma}}
\end{figure}

In order to quantify the many-body thermalization error, we define: 
\begin{equation}
\Delta=\sqrt{\frac{1}{N}\sum_{\nu=1}^{2^N}\left(G_{\nu}-G_{\nu}^{\rm thermal}\right)^{2}} \;.\label{eq:therm_error_nonint}
\end{equation}
In Fig (\ref{fig:therm_error_non_int_vs._gamma})
we present the thermalization error as a function of $\gamma$.   As seen in the figure, strict thermalization is only achieved for $\gamma\to 0$, and for small $\gamma$ the thermalization error $\Delta$  is linear in $\gamma$. The results are shown for several system sizes and are consistent with an error scaling as $N^{1/2}$ (see inset). This is the same behavior argued for in the integrable case, see Sec. \ref{sec:error}.

\section{Discussion}

The paper addresses thermalization of quantum system using a Lindblad approach. A new method was devised which allows the Lindblad operators to thermalize the system without coupling to the whole system. The method relies on acting on a lead with Lindblad jump operators and coupling the lead to the system, which is subjected to Hamiltonian dynamics only. For the method to be properly applied, the coupling of the system to the lead and of the Lindblad operators to the lead have to be weak. In these limits, the method ensures that there is both thermalization into a Gibbs state and that fluctuation-dissipation relations are satisfied.  The corresponding errors as a function of the coupling were estimated. The study also shows the limitations of applying the Lindblad approach to strongly driven systems.

 This approach can serve as the starting point to study transport in situations where the system is coupled to multiple leads.  The resulting behavior of the system is expected to be similar to that obtained by a Landauer approach for non-interacting systems and wealky interacting systems\cite{Landauer1957,Meir1992,Gruss2016}. However, the method can also be applied to systems of arbitrarily strong interactions, and gives access to the full density matrix of the system.
 
\acknowledgements{We are grateful to Martin Fraas, Andrea Gambassi, Misha Reznikov, and Ari Turner for useful discussions. This work was partially supported by the Israel Science Foundation (ISF)
grant numbers 1839/13 and 1331/17, by the Joint UGS-ISF Research Grant Program under grant number 1903/14, and  by the National Science Foundation through a grant to ITAMP at the Harvard-Smithsonian Center for Astrophysics and through support from the Harvard-MIT CUA.}

\appendix

\section{Single-Site Coupling}
\label{appen:onesite}

The unique steady state of the Lindblad equation, with the unitary part of the evolution given in Eq. (\ref{eq:one_site_hamiltonian}) and the dissipative generator in Eq. (\ref{eq:one_site_lindblad}), can be found by considering the following Ansatz,
\begin{eqnarray}
\tilde{\rho} & = & \left(g_{L}b^{\dagger}b+\left(1-g_{L}\right)bb^{\dagger}\right)\nonumber \\
 &  & \times\prod_{i=1}^{N}\left(g_{i}c_{i}^{\dagger}c_{i}+\left(1-g_{i}\right)c_{i}c_{i}^{\dagger}\right),\label{eq:one_site_lead_ansatz}
\end{eqnarray}
corresponding to a product density matrix in the site basis, with the parameters $g_L$ and $g_i$ to be determined. Substituting Eq. (\ref{eq:one_site_lead_ansatz}) into the dissipative generator we find,
\begin{eqnarray}
\hat{\Gamma}\tilde{\rho} & = & \gamma\commopen{f_{0}\left(1-g_{L}\right)\left(b^{\dagger}b-bb^{\dagger}\right)} +\commclose{\left(1-f_{0}\right)g_{L}\left(bb^{\dagger}-b^{\dagger}b\right)}\nonumber \\
 &  & \times\prod_{i=1}^{N}\left(g_{i}c_{i}^{\dagger}c_{i}+\left(1-g_{i}\right)c_{i}c_{i}^{\dagger}\right),
\end{eqnarray}
which vanishes for $g_L=f_0$.  Similarly, substituting into the unitary part of the evolution,
\begin{eqnarray}
\left[H,\tilde{\rho}\right] & = & t^{\prime}\left(g_{L}-g_{1}\right)\left(c_{1}^{\dagger}b-b^{\dagger}c_{1}\right)\nonumber \\
 &  & \times\prod_{j=2}^{N-1}\left(g_{j}c_{j}^{\dagger}c_{j}+\left(1-g_{j}\right)c_{j}c_{j}^{\dagger}\right)\nonumber \\
 &  & +\sum_{i=1}^{N}\left(g_{i}-g_{i+1}\right)\left(c_{i+1}^{\dagger}c_{i}-c_{i}^{\dagger}c_{i+1}\right)\nonumber \\
 &  & \times\prod_{j\neq i,i+1}^{N-1}\left(g_{j}c_{j}^{\dagger}c_{j}+\left(1-g_{j}\right)c_{j}c_{j}^{\dagger}\right) \nonumber \\
 &  & \times\left(g_{L}b^{\dagger}b+\left(1-g_{L}\right)bb^{\dagger}\right).
\end{eqnarray}
We thus see that for
\begin{equation}
g_{L}=g_{j}= f_0\qquad\forall j
\end{equation}
the unitary evolution $-i\left[H,\tilde{\rho}\right]=0$ also vanishes.

The analysis above shows that within this coupling protocol, different sites are uncorrelated in the steady state.  In fact, since the occupations are site-independent, it follows that the density matrix factorizes and is state independent for any one-particle basis. In particular, the reduced density matrix of the system (after tracing out the lead) becomes, in the energy basis,
\begin{eqnarray}
\rho_S=\prod_n \left(f_0 c_n^\dagger c_n+(1-f_0) c_n c_n^\dagger\right).
\end{eqnarray}
Hence, a state's occupation is seen to be independent of its energy, corresponding to an infinite temperature steady state, regardless of the value of $\beta$.  From the derivation above, it is clear that this result holds
for more general tight-binding models.

\section{Perturbation Theory\label{appen:PertTheory}}

In this Appendix, we consider general tight binding models. The setup is similar to Fig. \ref{fig:physical_setup}: 
\begin{equation}
H=H_{L}+H_{S}+H_{{\rm int}}
\end{equation}
Here, it is more convenient to work with the one-particle eigenstates of the lead and system Hamiltonians. Therefore, 
\begin{align}
H_{L} & =\sum_{m}\epsilon_{m}^{L}b_{m}^{\dagger}b_{m}\nonumber \\
H_{S} & =\sum_{n}\epsilon_{n}^{S}c_{n}^{\dagger}c_{n}\nonumber \\
H_{{\rm int}} & =\sum_{m,n}t_{mn}^{\prime}b_{m}^{\dagger}c_{n}+{\rm h.c.}
\end{align}
where $m$, $n$ are the eigenstates of the lead, system Hamiltonians, respectively, $b_{m}$, $c_{n}$ are their corresponding annihilation operators and $\epsilon_{m}^{L}$, $\epsilon_{n}^{S}$ the corresponding energies. Right now, we don't assume specific details about the Hamiltonians, except that $H_{{\rm int}}$ is governed by an overall scale $t^{\prime}$ and vanishes as $t^{\prime}\to0$.

Our goal is to obtain the solution to Eq. (\ref{eq:lindblad_steady_state})
order by order in $t^{\prime}$. In particular, we show that in the limit $t^{\prime}\rightarrow0$, the results in Sec. \ref{sec:weakCoupling} hold. 

We begin by denoting: 
\begin{equation}
\rho=\sum_{k=0}^{\infty}\rho_{k}\label{eq:rho_series}
\end{equation}
With $\rho_{k}\sim\left(t^{\prime}\right)^{k}$. Plugging Eq. (\ref{eq:rho_series}) into Eq. (\ref{eq:lindblad_steady_state}) and comparing terms of the same order, we can solve for the density matrix iteratively: 
\begin{align}
\left(\hat{H}_{0}+\hat{\Gamma}\right)\rho_{0} & = 0\label{eq:perturbation_theory_zeroth_order}\\
k\geq1:\quad\rho_{k} & = \left(\hat{H}_{0}+\hat{\Gamma}\right)^{-1}\left(-\hat{V}\right)\rho_{k-1}\label{eq:perturbation_theory_higher_orders}
\end{align}
Here, we focus on computations up to second order in $t^{\prime}$.
Higher order terms can be obtained iteratively, but are outside the scope of this paper.

Eq. (\ref{eq:perturbation_theory_zeroth_order}) can be satisfied by choosing a factorizable density matrix as follows: 
\begin{align}
\rho_{0} & = \prod_{m}\left(f_{m}b_{m}^{\dagger}b_{m}+\left(1-f_{m}\right)b_{m}b_{m}^{\dagger}\right)\nonumber \\
 &\times \prod_{n}\left(g_{n}c_{n}^{\dagger}c_{n}+\left(1-g_{n}\right)c_{n}c_{n}^{\dagger}\right)\label{eq:rhotp0_2}
\end{align}
with $0\leq g_{n}\leq1$, at this point arbitrary. Without loss of generality (see remark at the end of the Appendix), we assume that $H_{S}$ is non-degenerate. In this case, the above form of density matrix is the only possible choice to satisfy Eq. (\ref{eq:perturbation_theory_zeroth_order}). However, the subspace of solutions to this equation is largely degenerate. Therefore, to determine $\rho_{0}$, a degenerate perturbation theory has to be employed.

Note that here, in contrast to a usual perturbation theory, the Schrodinger equation is replaced by the Lindblad equation, and the Fock space by its dual, namely the space of the linear operators which act on
Fock states. Also, since we are interested in the steady state, the eigenvalue of $\hat{H}_{0}+\hat{\Gamma}+\hat{V}$ is $0$ to all orders
in $t^{\prime}$.

\begin{table}[!bh]
\centering \caption{Eigenstates of $\hat{H}_{0}+\hat{\Gamma}$}
\label{tab:righteigenstates} 
\begin{ruledtabular}
\begin{tabular}{@{}llr@{}}
Name  & Expression  & Eigenvalue \\
\hline 
$\Omega_{0,m}^{L}$  & $f_{m}b_{m}^{\dagger}b_{m}+\left(1-f_{m}\right)b_{m}b_{m}^{\dagger}$  & $0$ \\
$\Omega_{1,m}^{L}$  & $b_{m}^{\dagger}$  & $-i\epsilon_{m}^{L}-\frac{\gamma}{2}$ \\
$\Omega_{2,m}^{L}$  & $b_{m}$  & $i\epsilon_{m}^{L}-\frac{\gamma}{2}$ \\
$\Omega_{3,m}^{L}$  & $b_{m}^{\dagger}b_{m}-b_{m}b_{m}^{\dagger}$  & $-\gamma_{m}$ \\
$\Omega_{0,n}^{S}$  & $g_{n}c_{n}^{\dagger}c_{n}+\left(1-g_{n}\right)c_{n}c_{n}^{\dagger}$  & $0$ \\
$\Omega_{1,n}^{S}$  & $c_{n}^{\dagger}$  & $-i\epsilon_{n}^{s}$ \\
$\Omega_{2,n}^{S}$  & $c_{n}$  & $i\epsilon_{n}^{s}$ \\
$\Omega_{3,n}^{S}$  & $c_{n}^{\dagger}c_{n}-c_{n}c_{n}^{\dagger}$  & $0$ \\
\end{tabular}
\end{ruledtabular}
\end{table}

\begin{table}[!bh]
\centering \caption{Adjoints of the Eigenstates of $\hat{H}_{0}+\hat{\Gamma}$}
\label{tab:lefteigenstates} 
\begin{ruledtabular}
\begin{tabular}{@{}llr@{}}
Name  & Expression  & Eigenvalue \\
\hline 
$\Omega_{0,m}^{L,{\rm ad}}$  & $b_{m}^{\dagger}b_{m}+b_{m}b_{m}^{\dagger}=\mathbb{I}_{m}$  & $0$ \\
$\Omega_{1,m}^{L,{\rm ad}}$  & $b_{m}$  & $-i\epsilon_{m}^{L}-\frac{\gamma}{2}$ \\
$\Omega_{2,m}^{L,{\rm ad}}$  & $b_{m}^{\dagger}$  & $i\epsilon_{m}^{L}-\frac{\gamma}{2}$ \\
$\Omega_{3,m}^{L,{\rm ad}}$  & $\left(1-f_{m}\right)b_{m}^{\dagger}b_{m}-f_{m}b_{m}b_{m}^{\dagger}$  & $-\gamma_{m}$ \\
$\Omega_{0,n}^{S,{\rm ad}}$  & $c_{n}^{\dagger}c_{n}+c_{n}c_{n}^{\dagger}=\mathbb{I}_{n}$  & $0$ \\
$\Omega_{1,n}^{S,{\rm ad}}$  & $c_{n}$  & $-i\epsilon_{n}^{s}$ \\
$\Omega_{2,n}^{S,{\rm ad}}$  & $c_{n}^{\dagger}$  & $i\epsilon_{n}^{s}$ \\
$\Omega_{3,n}^{S,{\rm ad}}$  & $\left(1-g_{n}\right)c_{n}^{\dagger}c_{n}-g_{n}c_{n}c_{n}^{\dagger}$  & $0$ \\
\end{tabular}
\end{ruledtabular}
\end{table}

For convenience, we introduce some notation. Since the unperturbed operator $\hat{H}_{0}+\hat{\Gamma}$ is a sum of one-particle operators, its eigenstates can be factorized: 
\begin{align}
\rho_{\left\{ i_{m},i_{n}\right\} } & =\prod_{m}\Omega_{i_{m},m}^{L}\otimes\prod_{n}\Omega_{i_{n},n}^{S},
\end{align}
where $i_{m}$, $i_{n}$ are indices taking the values of $0-3$, and $\Omega_{i_{m},m}^{L}$ ($\Omega_{i_{n},n}^{S}$) are one-particle lead (system) eigenstates, respectively, specified in Table \ref{tab:righteigenstates}. The corresponding eigenvalue of $\rho_{\left\{ i_{m},i_{n}\right\} }$ is
\begin{align}
\left(\hat{H}_{0}+\hat{\Gamma}\right)\rho_{\left\{ i_{m},i_{n}\right\} } & =\rho_{\left\{ i_{m},i_{n}\right\} }\times\nonumber \\
 & \left(\sum_{m}\omega_{i_{m},m}^{L}+\sum_{n}\omega_{i_{n},n}^{S}\right)
\end{align}
with $\omega_{i_{m},m}^{L}$, $\omega_{i_{n},n}^{S}$ the one-particle eigenvalues (also specified in Table \ref{tab:righteigenstates}). Note that $\Omega_{0,n}^{S}$, $\Omega_{3,n}^{S}$ are degenerate, and we emphasize again that the $g_{n}$'s are arbitray. This apparently cumbersome choice of $\Omega_{0,n}^{S}$, $\Omega_{3,n}^{S}$ will prove to be useful later.

To formulate the perturbation theory in the dual space, we define the inner product between two states $\rho_{a}$ and $\rho_{b}$:
\begin{equation}
\left\langle \rho_{a},\rho_{b}\right\rangle \equiv{\rm Tr}\left(\rho_{a}^{{\rm ad}}\rho_{b}\right)
\end{equation}
However, since $\hat{H}_{0}+\hat{\Gamma}$ is not Hermitian, the adjoint state is not necessarily the Hermitian conjugate of the original state. Nevertheless, obtaining the adjoint state is relatively straightforward. The adjoint eigenstates $\rho_{\left\{ i_{m},i_{n}\right\} }^{{\rm ad}}$ are obtained by factorizing adjoint one-particle eigenstates, which are given in Table \ref{tab:lefteigenstates}. These, in turn, were obtained by requiring 
\begin{align}
\left\langle \Omega_{i_{m},m}^{L},\Omega_{j_{m},m}^{L}\right\rangle  & =\delta_{i_{m},j_{m}}\nonumber \\
\left\langle \Omega_{i_{n},n}^{S},\Omega_{j_{n},n}^{S}\right\rangle  & =\delta_{i_{n},j_{n}}
\end{align}
Employing the above notations, $\rho_{0}$ can be written simply as
\begin{equation}
\rho_{0}=\prod_{m}\Omega_{0,m}^{L}\prod_{n}\Omega_{0,n}^{S}
\end{equation}

We now proceed to find other solutions to Eq. (\ref{eq:perturbation_theory_zeroth_order}). These solutions form a subspace, which we name the ``ground state manifold", that is spanned by the following basis vectors:
\begin{equation}
\rho_{\left\{ i_{0,n}\right\} }=\prod_{m}\Omega_{0,m}^{L}\prod_{n}\Omega_{i_{0,n},n}^{S}\qquad i_{0,n}=0\;\text{or}\;3\label{eq:perturbation_theory_ground_state_manifold}
\end{equation}

The degenerate perturbation theory requires that $\rho_{1}$ (Eq. (\ref{eq:perturbation_theory_higher_orders})) has zero overlap with any $\rho_{\left\{ i_{0,n}\right\} }$. By applying $\hat{V}$ one obtains the following: 
\begin{align}
\left(-\hat{V}\right)\rho_{0} & =i\sum_{m,n}t_{mn}^{\prime}b_{m}^{\dagger}c_{n}\times\nonumber \\
 & \prod_{\mu\neq m}\Omega_{0,\mu}^{L}\prod_{\nu\neq n}\Omega_{0,\nu}^{S}\left(g_{n}-f_{m}\right)+{\rm h.c.}
\end{align}
and, trivially, $\rho_{1}$ has no overlap with the ground state manifold. Therefore, we proceed to second order in $t^{\prime}$: 
\begin{align}
\hat{V}\left(\hat{H}_{0}+\hat{\Gamma}\right)^{-1}\hat{V}\rho_{0} & =\prod_{\mu}\Omega_{0,\mu}^{L}\times\nonumber \\
 & \sum_{m,n}\frac{\left|t_{mn}^{\prime}\right|^{2}\gamma_{m}\left(g_{n}-f_{m}\right)}{\left(\frac{\gamma_{m}}{2}\right)^{2}+\left(\epsilon_{m}^{L}-\epsilon_{n}^{S}\right)^{2}}\times\nonumber \\
 & \left(\prod_{\nu\neq n}^{N}\Omega_{0,\nu}^{S}\right)\Omega_{3,n}^{S}\nonumber \\
 & +\rho_{2}^{\perp}\label{eq:perturbation_theory_rho_second_order}
\end{align}
with $\rho_{2}^{\perp}$ being terms outside the ground-state manifold. The requirement that $\left\langle \rho_{2},\rho_{\left\{ i_{0,n}\right\} }\right\rangle =0$ for any $\rho_{\left\{ i_{0,n}\right\} }$ is mostly trivial, except for the set of $N$ independent quations: 
\begin{equation}
\sum_{m}\frac{\left|t_{mn}^{\prime}\right|^{2}\gamma_{m}\left(g_{n}-f_{m}\right)}{\left(\frac{\gamma_{m}}{2}\right)^{2}+\left(\epsilon_{m}^{L}-\epsilon_{n}^{S}\right)^{2}}=0\qquad\forall\qquad n\label{eq:perturbation_theory_secular_equation}
\end{equation}
which results in Eq (\ref{eq:occupations_perturbation_theory}), as expected. In addition to the occupations $g_{n}$, we can obtain different correlation functions from $\rho_{2}^{\perp}$ (Eq. (\ref{eq:perturbation_theory_rho_second_order})), up to second order: 
\begin{align}
\left\langle b_{m}^{\dagger}b_{m^{\prime}}\right\rangle _{m\neq m^{\prime}}^{\left(2\right)} & = \frac{1}{-i\left(\epsilon_{m^{\prime}}^{L}-\epsilon_{m}^{L}\right)-\frac{\gamma_{m}+\gamma_{m^{\prime}}}{2}}\times\nonumber \\
  & \sum_{n}\frac{\left(t_{mn}^{\prime}\right)^{\star}t_{m^{\prime}n}^{\prime}\left(g_{n}-f_{m}\right)}{i\left(\epsilon_{m}^{L}-\epsilon_{n}^{S}\right)-\frac{\gamma_{m}}{2}}\nonumber \\
  & +{\rm c.c.},m\leftrightarrow m^{\prime}
\end{align}

\begin{align}
\left\langle b_{m}^{\dagger}b_{m}-b_{m}b_{m}^{\dagger}\right\rangle ^{\left(2\right)} & =  \frac{1}{-\gamma_{m}}\sum_{n}\frac{\left|t_{mn}^{\prime}\right|^{2}\left(g_{n}-f_{m}\right)}{-i\left(\epsilon_{m}^{L}-\epsilon_{n}^{S}\right)-\frac{\gamma_{m}}{2}}\nonumber \\
 &   +{\rm c.c.}
\end{align}

\begin{align}
\left\langle c_{n}^{\dagger}c_{n^{\prime}}\right\rangle _{n\neq n^{\prime}}^{\left(2\right)} & = \frac{1}{-i\left(\epsilon_{n^{\prime}}^{s}-\epsilon_{n}^{S}\right)}\times\nonumber \\
  & \sum_{m,n^{\prime\prime}}-\frac{t_{mn}^{\star}\left(t_{mn^{\prime}}^{\prime}\right)^{\star}\left(g_{n}-f_{m}\right)}{-i\left(\epsilon_{m}^{L}-\epsilon_{n}^{S}\right)-\frac{\gamma_{m}}{2}}\nonumber \\
  & +{\rm c.c.},n\leftrightarrow n^{\prime}
\end{align}

The lead-system correlations can be obtained as well, and can be useful in obtaining the current in out of equilibrium setups: 
\begin{equation}
\left\langle b_{m}^{\dagger}c_{n}\right\rangle ^{\left(1\right)}=\frac{i\left(t_{mn}^{\prime}\right)^{\star}\left(g_{n}-f_{m}\right)}{i\left(\epsilon_{m}^{L}-\epsilon_{n}^{s}\right)-\frac{\gamma_{m}}{2}}
\end{equation}
Note that these processes reflect the virtual creation and annihilation of the particles in the lead and system via the weak coupling. 

As a final remark, we would like to clarify that this treatment can be generalized to a degenerate $H_{S}$.  For example, in the simplest case, if the degeneracy results from known symmetries, $H_{S}$ can be diagonalized simultaneously with the additional conserved quantities, and then these eigenstates are chosen as the constituents of Eq. (\ref{eq:rhotp0_2}).

\bibliography{paper_updated}

\end{document}